\documentclass[a4paper,twoside]{article}
\usepackage[dvips]{graphicx}
\usepackage{latexsym,epsf}
\usepackage{multicol}
\usepackage[centertags]{amsmath}
\usepackage [cp1251]{inputenc}
\usepackage [ukrainian, russian, english]{babel}
\usepackage {fancyhdr}

\fancyfootoffset{1pt}%
\renewcommand {\footrule}{\vbox to 0pt{\hbox to \headwidth{ \hrulefill \hspace{63mm}}\vss}}

\makeatletter
\renewcommand{\ps@plain}{
\renewcommand{\@oddhead}{}
\renewcommand{\@evenhead}{}
\renewcommand{\@oddfoot}{\hfil \thepage}
\renewcommand{\@evenfoot}{\thepage \hfil \hfil}}
\makeatother 

\makeatletter
\renewcommand{\@biblabel}[1]{#1.\hfill}
\makeatother
\title{\textbf{\Large COMPARISON OF $GEANT4$ WITH $EGSnrc$ FOR SIMULATION OF GAMMA-RADIATION DETECTORS BASED ON SEMI-INSULATING MATERIALS}}
\author{\textbf{\textit{A.I. Skrypnyk\footnote{\normalfont
Corresponding author E-mail address: belkas@kipt.kharkov.ua}, A.A. Zakharchenko, M.A. Khazhmuradov
}}
\\
\emph{\small National Science Center "Kharkov Institute of
Physics and Technology", 61108,  Kharkov, Ukraine}\\
{\small(Received August 1 , 2011)}}
\begin{document}
\selectlanguage{english}
\date{}
\maketitle

\thispagestyle{fancy}

\begin{center}
\begin{minipage}{165mm}
{\small
We considered $GEANT4$ version 4.9.4 with different Electromagnetic Physics Package for calculation of response functions of detectors based on semi-insulating materials. Computer simulations with $GEANT4$ packages were run in order to determine the energy deposition of gamma-quanta in detectors of specified composition (HgI$_2$ and TlBr) at various energies from 0.026 to 3~MeV. The uncertainty in these predictions is estimated by comparison of their results with $EGSnrc$ simulations. A general good agreement is found for $EGSnrc$ and $GEANT4$ with $Penelope$ 2008 model of LowEnergy Electromagnetic package.}
\par \vspace{1ex}
\end{minipage}
\end{center}
\begin{multicols}{2}
\begin{center}
\textbf{\textsc{1. INTRODUCTION}}
\end{center}
The variety of electrophysical characteristics (specific resistance,
product of mobility ${\mu}$  and mean drift time ${\tau}$  for holes
and electrons -- (${\mu}{\tau}$)$_{h,e}$) is a serious
disadvantage of semi-insulating materials (wide band-gap semiconductor
with high resistivity). Now this is a main cause that the
semi-insulator gamma-radiation detectors could not be mass produced.
Wide band-gap semiconductor detectors have considerable spread of
(${\mu}{\tau}$)$_{h,e }$ values (in order of magnitude and
more) [1] even if they are produced from one ingot. As result the same
size detectors biased to the same voltage \textit{U}$_{b}$
have a different charge collection efficiency (\textit{CCE}). It results in non-uniformity of serial devices response and in necessity of individual setup parameters selection.

Also the material non-uniformities are a serious obstacle to comparison
of detector characteristics measured in different studies. Computer
simulation is an optimal method to overcome the material non-uniformity
problem. Simulation may be useful both to research of wide band-gap
detector behavior and designing of device based on them [2].

Previously we have developed the model of the planar wide band-gap
semiconducting gamma-radiation detectors where EGSnrc Monte-Carlo (MC)
simulation package was used for determination of the energy deposition of
gamma-quanta and charged particles [3]. This model was tested on the
several groups of $CdTe$ and $CdZnTe$ detectors [4]. The observed
difference between model and experimentally measured amplitude
distributions of radiation sources $^{137}$Cs and
$^{152}$Eu was explained in process of this model
verification [2]. However, by this now all factors defining such
important gamma-radiation detector characteristics as sensitivity
${\delta}$  and charge collection efficiency \textit{CCE} are not
determined even for the most researched semi-insulating materials as
$CdTe$ and $CdZnTe$. By-turn it does not permit to define exhaustive set of
the control parameters of the wide band-gap detector model. The
determination of these parameters would allow calculating the most
realistic response of the gamma-radiation detector.

The model [4] enabled to obtain a good agreement between
calculated and experimental response functions of $CdTe$ ($CdZnTe$)
gamma-radiation detectors in the most cases analyzed. However $EGSnrc$ package possibilities are
restricted only by a simulation of photon, electron and positron
transport. It does not allow using the above mentioned model [4] for
analysis of the charge collection efficiency experiments with the wide
band-gap detectors when proton and ${\alpha}$-particle beams are used
[5]. Also the model [4] is unsuitable for researching of actual
problems of radiation resistance of semi-insulators in mixed radiation
fields (charged particles and/or neutrons and gamma-quanta) [6].

$Geant4$ [7] is an actively developing toolkit for the simulation of the
passage of charged particles, neutrons and gamma-quanta through matter.
This MC package offers a set of physical process models to describe the
interaction of charged and neutral particles with matter in the wide
energy range. $Geant4$ provides various models of the same
electromagnetic processes (EM packages) [8]. To implement the model [4]
on $Geant4$ platform properly it is necessary to define EM package which
provides better agreement with $EGSnrc$ simulation results.

In the present work statistical characteristics of the $EGSnrc$-calculated
response functions of planar wide band-gap detectors based on
HgI$_2$ and TlBr to gamma-quanta with energies between 0.026 and 3~MeV is compared with the results of $Geant4$ simulation. The same radiation source and detector geometry parameters are used to calculate the photon and charged particle transport in $EGSnrc$ and $Geant4$ simulations. A good agreement between results of $EGSnrc$ and $Geant4$ simulations is obtained with using $Penelope$-2008 EM package $Geant4$ v.4.9.4 [9].

\begin{center}
\textbf{\textsc{2. THE $GEANT4$ AND $EGSnrc$ ELECTROMAGNETIC MODELS}}
\end{center}
$Geant4$ includes three EM packages, which use different models for cross-sections of photons and charged particles interactions with the matter and different models for final state sampling algorithms. These are Standard
EM model and two models for low energy region identified as $Livermore$
and $Penelope$ EM model. All models describe the photoelectric effect,
the Compton scattering and the gamma conversion. The model processes of
the induced emission and scattering e$^-$--e$^-$ and e$^-$--e$^+$ are available for electrons and positrons.

In the recent years many articles were published where $Geant4$
simulation results are compared with yield of  {\guillemotleft}reference{\guillemotright} MC codes $EGSnrc$ and $MCNP$
[7--10]. $EGSnrc$ and $MCNP$ packages have longer application history in
radiation physics and nuclear medicine in comparison with Geant4. As a
result of often modifications of $Geant4$ code and fixation of algorithm
bugs the comparison of different $Geant4$ version simulations with other
Monte-Carlo codes and experimental data can show the differing results.
In the present work $Geant4$ v.4.9.3 and v.4.9.4 simulations are compared
with results of $EGSnrc$ version 4r2.3.1.

The Standard model uses an analytical approach to describe the
electromagnetic interactions in the range from 1~keV up to about
100~TeV. An analytical approach combines numerical databases with
analytical cross-section models assuming quasi-free atomic electrons
while the atomic nucleus is fixed. Transport of X- and gamma rays
takes into account Compton scattering using the free-electron
approximation, gamma conversion into electron-positron pair, and
photoelectric effect. Bremsstrahlung and ionization are the available
processes for electrons and positrons. The binding energy of atomic
electrons is taken into account only at the photoelectric effect
simulation. The simplified computational algorithm of the nuclear
radiation transport through the matter provides the most calculating
efficiency for the Standard model in comparison with other models.
However, coherent (Rayleigh) scattering and atomic relaxation processes
are not included for this package.

Atomic and shell ionization effects are included in the $Livermore$ model.
The lower energy threshold of the simulation interactions is decreased
down to 250~eV. The available physical processes are Rayleigh and
Compton scattering, photo-electric effect, pair production,
bremsstrahlung, and ionization. The fluorescence and Auger emission in
excited atoms are also considered. This package describes the
electromagnetic interactions of electrons and photons taking into
account subshell integrated cross sections for photoelectric effect,
ionization, and electron binding energies for all subshells.
Cross-sections of particle interactions with matter are calculated from
evaluated data libraries from Lawrence Livermore National Laboratory --
EPDL97 for photons [11], EEDL for electrons [12] and EADL for
fluorescence and Auger effects [13].

The physical model which originally was developed for $Penelope$ MC code
[9] combines analytical approach for computation of cross-section of
the different interactions with data from cross-section databases
calculated in Seltzer and Berger work [14]. Algorithm $Penelope$ 2008 is
used in $Geant4$ beginning with version v.4.9.3. This model is applicable
to interactions in range from about 200~eV up to 1~GeV. Now $Penelope$ EM
Package is still tested.

NIST data [14, 15] with empirical correction factors for set of elements
[16] are employed for computation of bremsstrahlung loss in the $EGSnrc$
model of the electromagnetic interactions. Cross-sections of the
photoelectric effect, pair and triplet generation and coherent
scattering cross-sections are determined from EDPL [11] and XCOM [17]
estimations. The Klein-Nishina method allowing for Doppler broadening
correction and correction for chemical bond effects is used to
calculate the incoherent (Compton) scattering cross-sections.
Parameters of atom relaxation are determined for every atomic subshell.

Contribution of the atom relaxation processes can be appreciable during
registration of gamma-quanta with energies up to about 1~MeV. They are
not described in the $Geant4$ Standard Electromagnetic Model. Thus it is
possible to expect the largest difference between simulation data
received with using Standard EM Package and $EGSnrc$ simulation results.

\begin{center}
\textbf{\textsc{3. RESULTS OF SIMULATIONS}}
\end{center}
Functions of gamma-quantum energy loss distribution (response functions)
were computed for planar detectors based on two wide band-gap semi-conducting
compounds: mercuric (II) iodide (HgI$_{2}$) and thallium
bromide (TlBr). HgI$_{2}$ detectors are investigated over a
long period of time [18] while TlBr is a comparatively new material for
nuclear radiation detection [19]. Initial energy of gamma-quantum beam
was chosen in the range from 0.026 up to 3~MeV. Simulations were
run for all three $Geant4$ packages of electromagnetic interactions.
Geometric sizes of detectors and gamma-quantum beam parameters were
identical for all $Geant4$ and $EGSnrc$ models.

Since the thickness of investigated detectors was about 1~mm the minimal
run length of $Geant4$-simulated particles (\textit{cutForElectron}) and
gamma-quanta (\textit{cutForGamma}) in detector material was set equal
to 0.01 mm. It corresponds to the minimal energy 3.21~keV for
gamma-quanta and 43.23~keV for electrons in HgI$_{2}$. The
minimal energy of $Geant4$-simulated gamma-quanta in TlBr is 3.85~keV
and 48.14~keV for electrons. The similar energy thresholds
\textit{PCUT} = 2~keV (gamma-quanta) and \textit{ECUT} = 20~keV
(electrons) were set for both materials during $EGSnrc$ simulations.

In the present work we do not consider the energy loss of fast electrons
due to the excitation of lattice vibrations. Statistical moments: average energy loss, variance and asymmetry coefficient were computed for every function of distribution of energy loss. Received dependences of the statistical moments against initial energy of gamma-quantum beam allow revealing and describing in detail systematical differences between results of $Geant4$ and $EGSnrc$ simulation.

\begin{center}
\textbf{\textsc{3.1  Mercuric (II) Iodide}}
\end{center}
Mercuric (II) iodine (HgI$_2$) belongs to the group of A$^{II}$B$^{VI}$ semiconductor compounds. The functions of distribution of gamma-quantum energy loss were computed for the planar HgI$_2$ detector with size of
5$\times$5$\times$1~mm$^3$. The parallel monoenergetic gamma-quantum beam was uniformly scanned on the detector area. The beam incidence angle was equal to 90\textdegree{}. At the first stage of the pesent work we used $Geant4$ v.4.9.3-p01 for comparison with $EGSnrc$ simulations. 10$^6$ gamma-quantum histories were simulated for every energy. The statistical moments of the response functions were
calculated on their base.

Fig.~1 shows the dependence of average energy losses
{\textless}\textit{E}{\textgreater} of gamma-quanta (the first moment)
against initial beam energy in the investigated HgI$_2$
detector. Calculated energy losses are practically equal for all
investigated models in the energy range less than
\textit{E}$_{\gamma}$ {\textless} 0.4~MeV. The energy
losses calculated from Standard and Livermore simulation data are
stably lower than results of EGSnrc simulations in region
\textit{E}$_{\gamma}$ {\textgreater} 0.4~MeV.
\begin{minipage}{80 mm}
\includegraphics{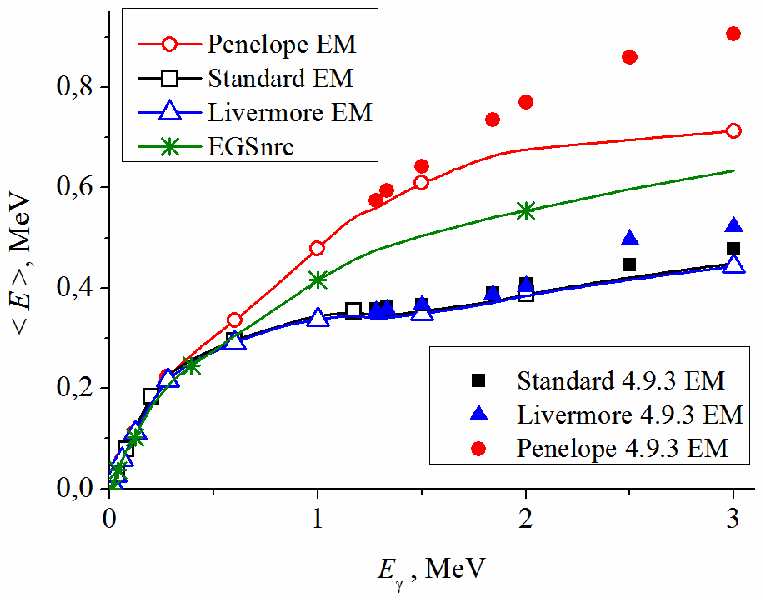}
\emph{\textbf{Fig.1.}} {\emph{The dependence of average energy losses of gamma-quanta in HgI$_2$ vs initial beam energy}}\\
\end{minipage}

The predicted mean energy losses of gamma-quanta received from Standard
and $Livermore$ models practically coincides in the range between 0.4 and
2~MeV. But the Geant4 v.4.9.3-p01 simulation demonstrates abrupt
increase of the mean losses for $Livermore$ model relative to Standard
model data for gamma-quantum energies above \textit{E}$_{\gamma}$ {\textgreater} 2~MeV (Fig.~1). The quick increase of the electron-positron pair production cross-section is a single essential feature of the gamma-quantum interaction with HgI$_2$ in energy range above
\textit{E}$_{\gamma}$ {\textgreater} 2~MeV. Consequently observed
difference (Fig.~1) may be concerned with $Geant4$ v.4.9.3-p01 errors of
the positron trajectory simulation.

The repeated simulation of the HgI$_2$-detector response
functions using $Geant4$ v.4.9.4-p02 (release from 24-06-2011) showed
that these bugs have been fixed by authors. Moreover, as Fig.~1 shows
the difference between $Penelope$ model data in version 4.9.4-p02 and
result of $EGSnrc$ simulation considerably decreased. That is why, except
where noted, for further study we use results received from $Geant4$
v.4.9.4-p02 simulations.

To reveal the most essential differences between simulation results
received with different $Geant4$ EM packages it is necessary the detailed
analysis of the investigated HgI$_2$-detector response
functions in the energy range of gamma-quanta where different physical
processes are dominated.

Fig.~2 shows the computed distributions of gamma-quantum energy losses
for initial beam energy \textit{E}$_{\gamma}$ = 0.08~MeV. At this energy the photoelectric effect is a dominating process of the gamma-radiation interaction with HgI$_2$. The amount of absorbed photons (on the Fig.~2 and further these events correspond to energy \textit{E}$_{\gamma}$) for all models insignificantly differs (Table~1). It confirms that similar
values of photoelectric effect cross-sections for Hg and I are used in
all models. Atomic relaxation processes are not taken into account in
the Standard model. Therefore the photoelectric effect event quantity
is maximal for this model.
\begin{minipage}{80 mm}
\includegraphics[width=\textwidth]{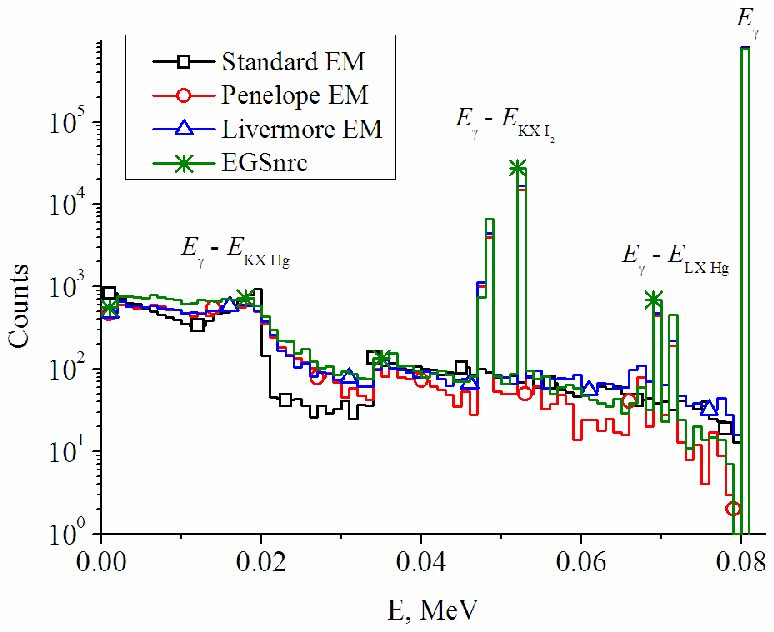}
\emph{\textbf{Fig.2.}} {\emph{Energy losses of gamma-quanta with energy
E$_{\gamma}$ = 0.08~MeV in HgI$_2$-detector}}\\
\end{minipage}

Table 1 and follow-up data show that the total efficiency of registration of gamma-quanta of investigated detectors is almost the same for all models.

The fine structure of escape peaks is reconstructed from the results of
$EGSnrc$ simulations in the most detail. As Fig.~2 demonstrates the used
energy bin (1~keV) allows separating \textit{KX}- and \textit{LX}-series of characteristic photons for both elements (Hg and I).

\begin{center}
\emph{\textbf{Table 1.}} {\it The amount of photoelectric effect events and total efficiency for gamma-quanta with \\ E\textup{${_\gamma}$} =
0.08~MeV}
\end{center}
\begin{center}
\begin{tabular}{|c|c|c|}
\hline
EM model & Photoeffect & Efficiency, HgI$_2$\\
\hline
Standard & \centering 804176 & 0.818\\
\hline
Livermore & \centering 784069 & 0.822\\
\hline
Penelope & \centering 788043 & 0.822\\
\hline
EGSnrc & \centering 765231 & 0.819\\\hline
\end{tabular}
\end{center}

In low-loss energy region of HgI$_2$-detector response functions computed with using $Livermore$ and $Penelope$ EM models show good agreement with $EGSnrc$ simulation data. According to Fig.~2 these three models have an almost equal distribution of the energy losses in the region up to 0.04{\dots}0.045~MeV. In the energy loss region above 0.045~MeV the better agreement with $EGSnrc$ data is observed for $Livermore$ EM model.

All EM models lead to the similar distribution of the Compton tail in
the gamma-quantum energy region where Compton scattering is prevailed
(Fig.~ 3). Computations with $Penelope$ EM model allow obtaining the
better agreement with $EGSnrc$ simulation in the Compton valley whereas
Standard and $Livermore$ results exceed $EGSnrc$ data. At the same time the
amount of absorbed gamma-quanta for the $Livermore$ model is equal about
two thirds of the same as other models (Table~2).
\begin{minipage}{80 mm}
\includegraphics[width=\textwidth]{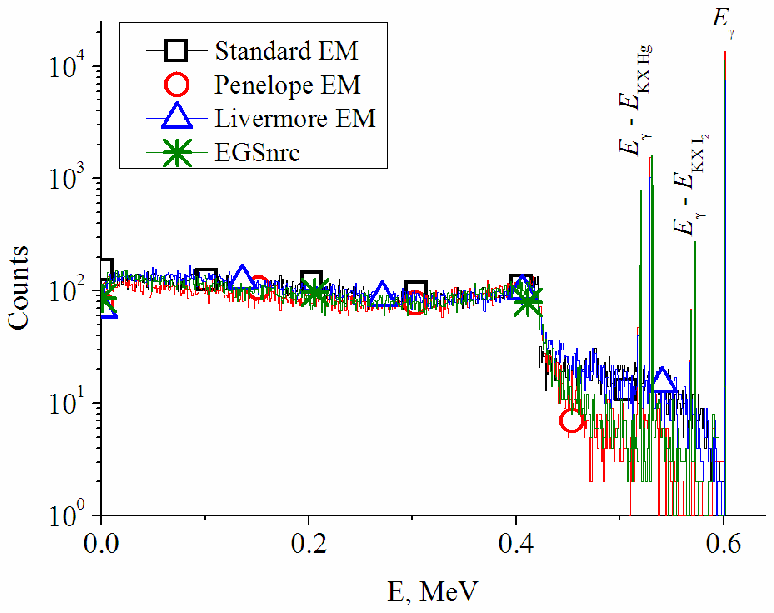}
\emph{\textbf{Fig.3.}} {\emph{Energy losses of gamma-quanta with energy
E$_{\gamma}$ = 0.6~MeV in HgI$_2$-detector}}\\
\end{minipage}

As follows Fig.~4 and Table~3 data the $Livermore$ and Standard EM models
employ cross-sections of photoelectric interaction with HgI$_2$ for gamma-quanta with energy \textit{E}$_{\gamma}$ = 1.28~MeV which values are much less compared with the same in the $Penelope$ EM model and $EGSnrc$. It is confirmed by the lack of photopeaks for the response functions simulated with using these models. Moreover the disappearing of the Compton edges on the Fig.~4 indicates that the $Livermore$ and Standard models employ smaller probabilities of the gamma-quantum Compton scattering angels above 135\textdegree{} in comparison with the $Penelope$ model and $EGSnrc$.

\begin{center}
\emph{\textbf{Table 2.}} {\it The amount of photoelectric effect events and total efficiency for gamma-quanta with \\ E\textup{${_\gamma}$} =
0.6~MeV}
\end{center}
\begin{center}
\begin{tabular}{|c|c|c|}
\hline
EM model & Photoeffect & Efficiency, HgI$_2$\\
\hline
Standard & 12602 & 0.058\\
\hline
Livermore & 7418 & 0.057\\
\hline
Penelope & 13488 & 0.058\\
\hline
EGSnrc & 11312 & 0.058\\
\hline
\end{tabular}
\end{center}

If gamma-quantum energy is enough for electron-positron pair production
(gamma-conversion) (for example \textit{E}$_{\gamma}$ =
1.28~MeV, Fig.~4) a double-escape peak of annihilation gamma-quanta becomes appreciable in the Geant4 simulations (Table~3). On the Fig.~4 these events correspond to the energy equal to 0.259~MeV. According to $EGSnrc$ simulation of response function of investigated HgI$_2$-detector at energy \textit{E}$_{\gamma}$ = 1.28~MeV the double escape peak is not yet observed (Table 3).
\begin{minipage}{80 mm}
\includegraphics[width=\textwidth]{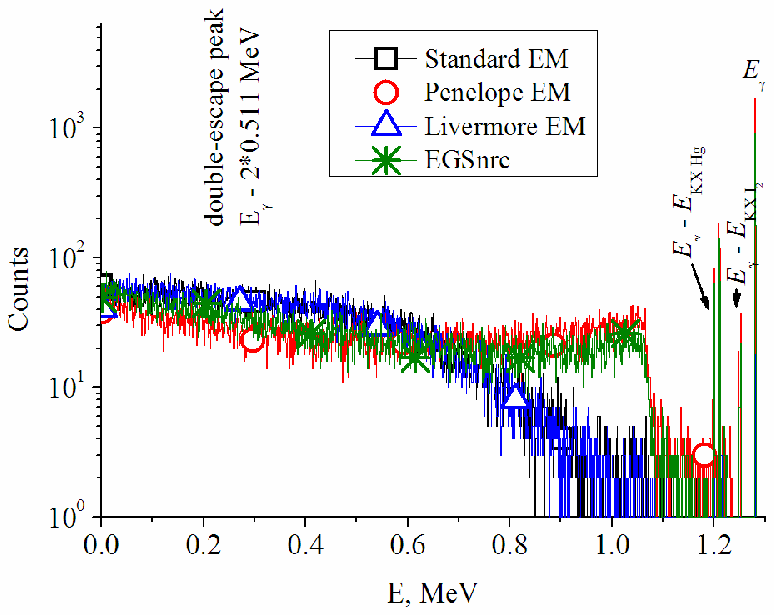}
\emph{\textbf{Fig.4.}} {\emph{Energy losses of gamma-quanta with energy E$_{\gamma}$ = 1.28~MeV in HgI$_2$-detector}}\\
\end{minipage}

\begin{center}
\emph{\textbf{Table 3.}} {\it The amount of the photoelectric effect and double escape events for gamma-quanta with \\ E\textup{$_{\gamma}$} = 1.28~MeV (HgI$_2$)}
\end{center}
\begin{center}
\begin{tabular}{|c|c|c|}
\hline
EM model & Photoeffect & Double-escape\\
\hline
Standard & 39 & 234\\
\hline
Livermore & 18 & 209\\
\hline
Penelope & 1700 & 201\\
\hline
EGSnrc & 1083 & 37\\
\hline
\end{tabular}
\end{center}

Above results show that there is good agreement between the response
functions of the investigated HgI$_2$-detector received with the $Penelope$ and $EGSnrc$ model apart from double-escape peak of the annihilation gamma-quanta.

Finally, the computation of the HgI$_2$-detector response function based on the 10$^7$ simulation trajectories of gamma-quanta with energy \textit{E}$_{\gamma}$ = 3~MeV shows a good agreement between results of the $Penelope$ EM model and $EGSnrc$ (Fig.~5). Plots corresponding to
results of the $Livermore$ and Standard models are located appreciably
lower. The escape peaks in that models are absent (Table~4 and Fig.~5). This indicates that values of cross-section of the electron-positron pair production employed by the $Livermore$ and Standard models are smaller as compared with the same of the $Penelope$ EM model and $EGSnrc$.
\begin{minipage}{80 mm}
\includegraphics[width=\textwidth]{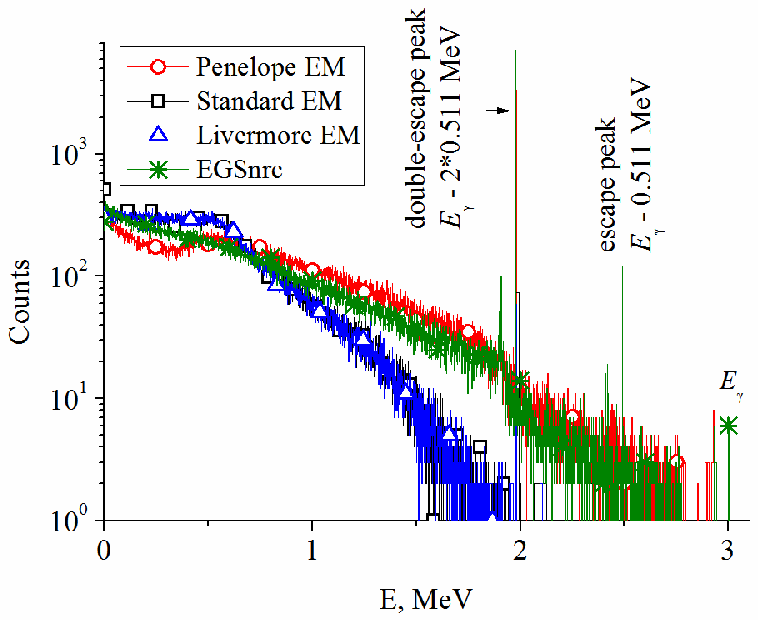}
\emph{\textbf{Fig.5.}} {\emph{Energy losses of gamma-quanta with energy E$_{\gamma}$ = 3~MeV in HgI$_2$-detector}}\\
\end{minipage}
\begin{center}
\emph{\textbf{Table 4.}} {\it The amount of event of the escape and double escape of annihilation gamma-quanta (HgI$_2$),
E{$_{\gamma}$} = 3~ MeV}
\end{center}
\begin{center}
\begin{tabular}{|c|c|c|}
\hline
EM model & Escape & Double-escape\\
\hline
Standard & 1 & 74\\
\hline
Livermore & 0 & 59\\
\hline
Penelope & 65 & 3366\\
\hline
EGSnrc & 121 & 7152\\
\hline
\end{tabular}
\end{center}

Fig.~5 shows that the response functions of the HgI$_2$-detector received with the $Geant4$ v4.9.4-p02 $Livermore$ and Standard models practically coincide. So average energy losses of gamma-quanta with energy 3~MeV also coincide (Fig.~1). It confirms our supposition about the incorrect computation of the HgI$_2$-detector response functions when $Geant4$ v.4.9.3-p01 $Livermore$ and Standard models have been used (Fig.~1).

Other statistical characteristics of response functions of the investigated HgI$_2$-detector -- variance (Fig.~6) and skewness coefficient (Fig.~7) -- also demonstrate a good agreement between results of $Geant4$ v4.9.4-p02 $Penelope$ EM package simulations and $EGSnrc$ data.

From Fig.~1, Fig.~6 and Fig.~7 it follows that main differences of all
statistical characteristics of the HgI$_2$-detector response functions between $Geant4$ v4.9.3-p01 and v4.9.4-p02 simulation data correspond to the electron-positron pair production region (\textit{E}$_{\gamma}$ {\textgreater} 1.022~MeV). In the gamma-quantum energy region
where positrons are not generated the response functions received with
different Geant4 versions coincide. Therefore the cause of
anomalous simulation results is incorrect values of specific
ionization energy losses of positrons in HgI$_2$ those used by $Geant4$ v4.9.3-p01. The differences between response functions of the investigated TlBr-detector received using of both $Geant4$ versions are insignificant.
\begin{minipage}{80 mm}
\includegraphics[width=\textwidth]{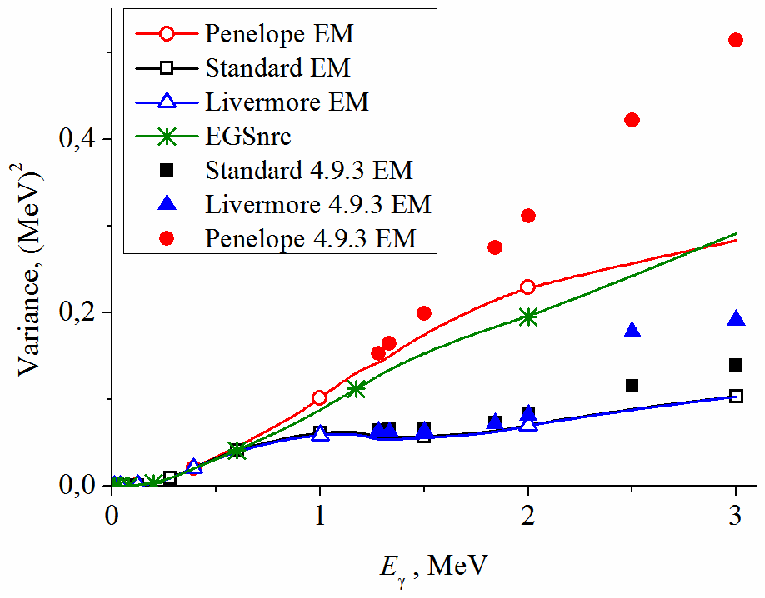}
\emph{\textbf{Fig.6.}} {\emph{The dependence of variance of the HgI$_2$-detector response functions vs gamma-quantum energy}}\\
\end{minipage}
\begin{minipage}{80 mm}
\includegraphics[width=\textwidth]{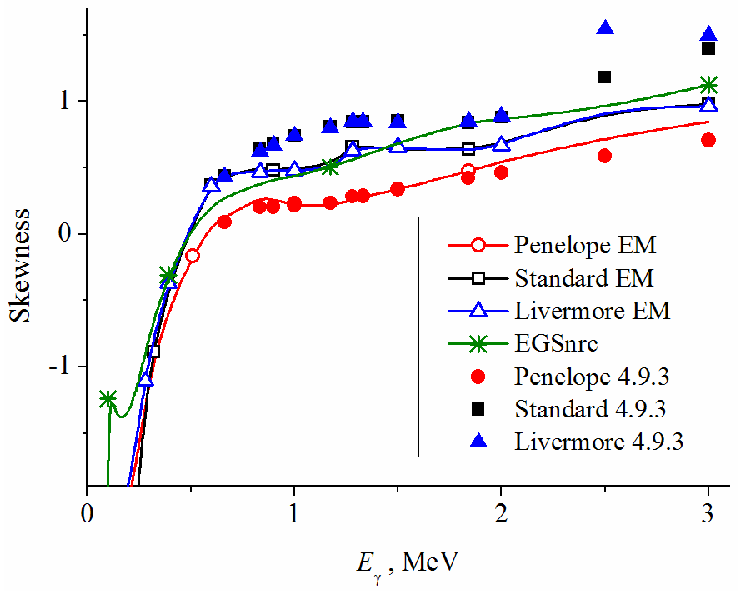}
\emph{\textbf{Fig.7.}} {\emph{The dependence of skewness coefficient of the HgI$_2$-detector response functions vs gamma-quantum
energy}}\\
\end{minipage}
\begin{center}
\textbf{\textsc{3.2 Thallium Bromide}}
\end{center}
Thallium bromide (TlBr) belongs to the group of A$^{III}$B$^{VI}$ semiconductor compounds. Quality differences between the linear gamma-ray attenuation coefficients of TlBr and HgI$_2$ are negligible. These
differences are characteristic for the energy region where photoelectric effect is dominated (Fig.~8). At the same time the TlBr linear gamma-ray attenuation coefficient exceeds the same of HgI$_2$ about 10\% beginning from gamma-ray energy \textit{E}$_{\gamma}$ = 0.1~MeV. As follows we can expect that average losses of gamma-quanta energy in TlBr will be
higher than in HgI$_2$ at the same detector thickness.

Distributions of the gamma-quantum energy losses were calculated for the planar TlBr-detector with size of 3.142~mm$^2$ $\times$ 0.8~mm. The parallel monoenergetic beam of gamma-quanta was uniformly distributed on the detector area. The beam incidence angle was equal to 90\textdegree{}. 10$^6$ histories were simulated for each gamma-ray energy. Statistical moments of response functions were calculated using these histories (Fig. 9 and Fig. 10).

The data plotted on the Fig.~9 -- Fig.~12 are received from the $Geant4$
v4.9.4-p02 simulations. Response functions of the investigated
TlBr-detector obtained using version v.4.9.3 coincide with version
v.4.9.4 calculations within the scope of statistical uncertainties. It
gives grounds for affirmation that the reason of above-mentioned
anomalous simulation results for HgI$_2$-detector is not
the errors of the positron trajectory simulation algorithm. As we think
these anomalous results are consequence of incorrect values of the
specific ionization energy losses of positrons in HgI$_2$.
Similarly the case of HgI$_2$-detector the better coincidence between $EGSnrc$ and $Geant4$ simulation data is observed when the $Penelope$ model of electromagnetic interactions has been used (Fig.~9 and Fig.~10).
\begin{minipage}{80 mm}
\includegraphics[width=\textwidth]{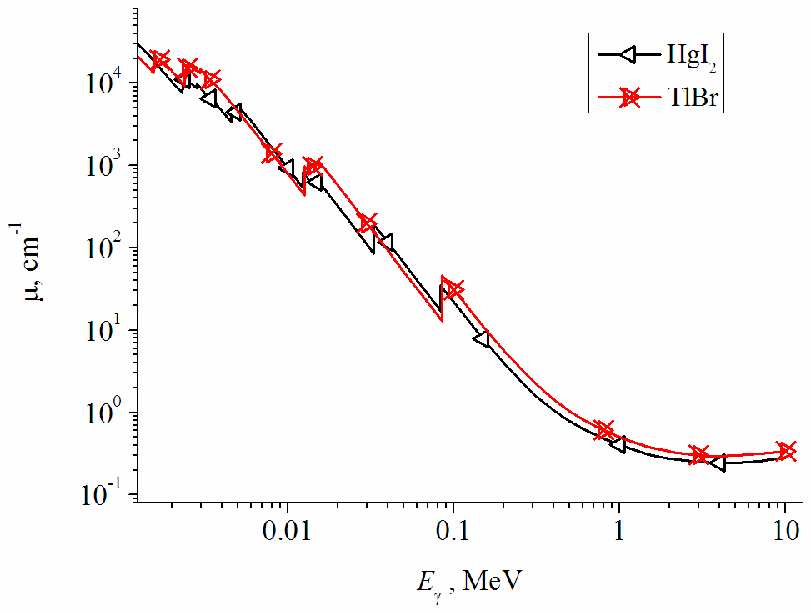}
\emph{\textbf{Fig.8.}} {\emph{The linear gamma-ray attenuation coefficient for HgI$_2$ and TlBr [17]}}\\
\end{minipage}

The analysis of simulated TlBr-detector response functions for gamma-ray
photons with energy \textit{E}$_{\gamma}$ = 0.6~MeV (Fig. 11) shows the presence of escape peaks of characteristic photons of Tl and Br $KX$-series in all models apart from the Standard EM model. The Compton tail is identically reconstructed in all simulations. As for the Compton valley region that the better agreement with $EGSnrc$ data is obtained for the $Geant4$
$Penelope$ EM package simulation. The amount of the absorbed gamma-quanta with the energy 0.6~MeV for the $Livermore$ EM model simulation is about 30\% less in comparison with other models (Table~5).

The response function of the investigated TlBr-detector for rather high
energy gamma-quanta (\textit{E}$_{\gamma}$ = 3~MeV, Fig. 12) received with using $EGSnrc$ MC package shows a good agreement only with the results of the $Geant4$ $Penelope$ model simulation. For the $Livermore$ and Standard models the escape peak of the annihilation gamma-quanta is almost absent and amplitude of the double escape peak is more than an order of magnitude less as compared with other simulations (Table~6).
\begin{minipage}{80 mm}
\includegraphics[width=\textwidth]{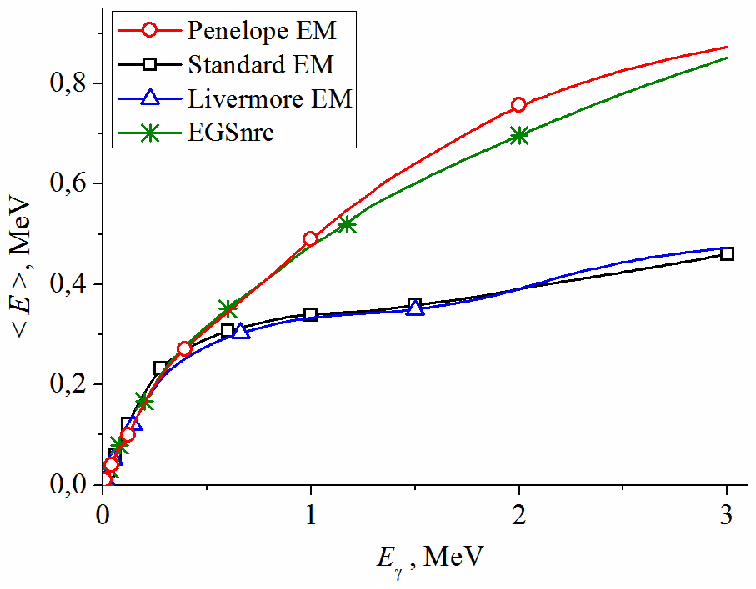}
\emph{\textbf{Fig.9.}} {\emph{The dependence of average energy losses of gamma-quanta in TlBr vs the initial energy beam}}\\
\end{minipage}
\begin{minipage}{80 mm}
\includegraphics[width=\textwidth]{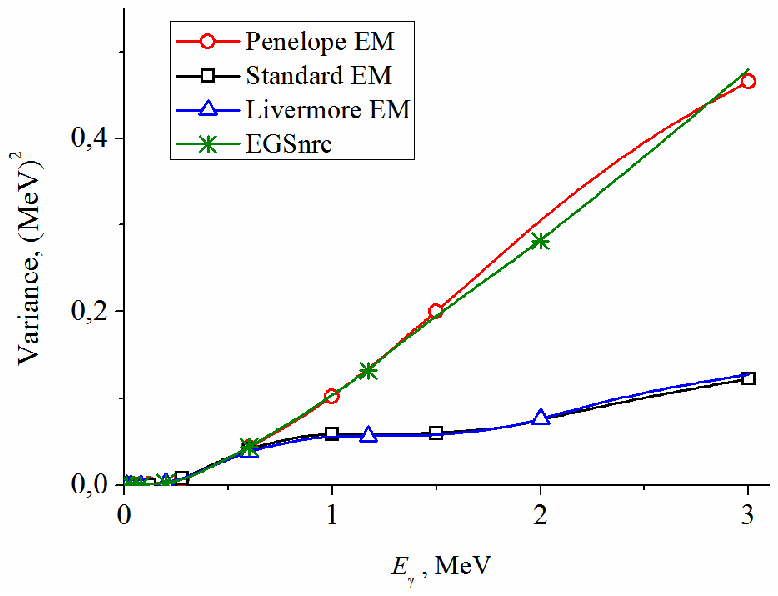}
\emph{\textbf{Fig.10.}} {\emph{The dependence of variance of TlBr-detector response functions vs the gamma-quantum energy}}\\
\end{minipage}
\begin{minipage}{80 mm}
\includegraphics[width=\textwidth]{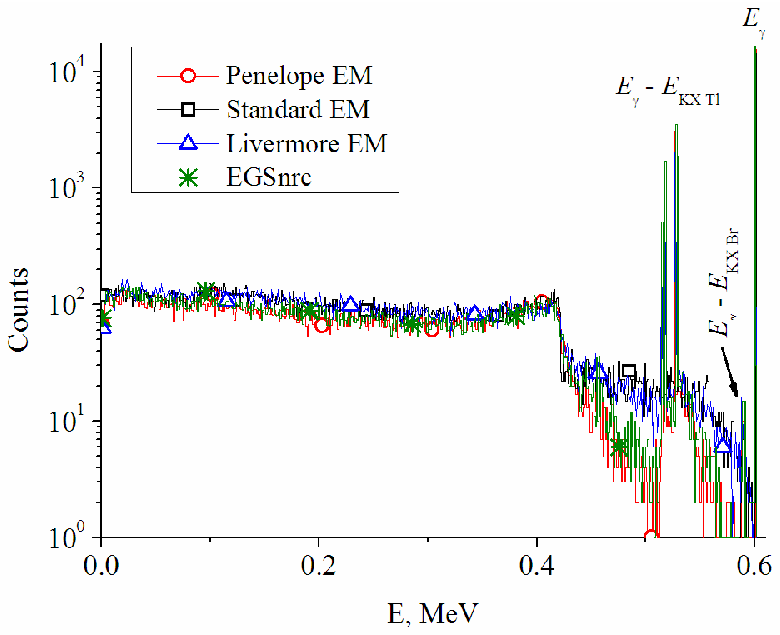}
\emph{\textbf{Fig.11.}} {\emph{The energy losses of gamma-quanta with the energy E$_{\gamma}$ = 0.6~MeV in TlBr-detector}}\\
\end{minipage}

\begin{center}
\emph{\textbf{Table 5.}} {\it The amount of photoelectric effect events and total efficiency for gamma-quanta with \\ E\textup{${_\gamma}$} =
0.6~MeV}
\end{center}
\begin{center}
\begin{tabular}{|c|c|c|}
\hline
EM model & Photoeffect & Efficiency, TlBr\\
\hline
Standard & 14288 & 0.061\\
\hline
Livermore & 9231 & 0.060\\
\hline
Penelope & 15652 & 0.060\\
\hline
EGSnrc & 16804 & 0.064\\
\hline
\end{tabular}
\end{center}
\begin{minipage}{80 mm}
\includegraphics[width=\textwidth]{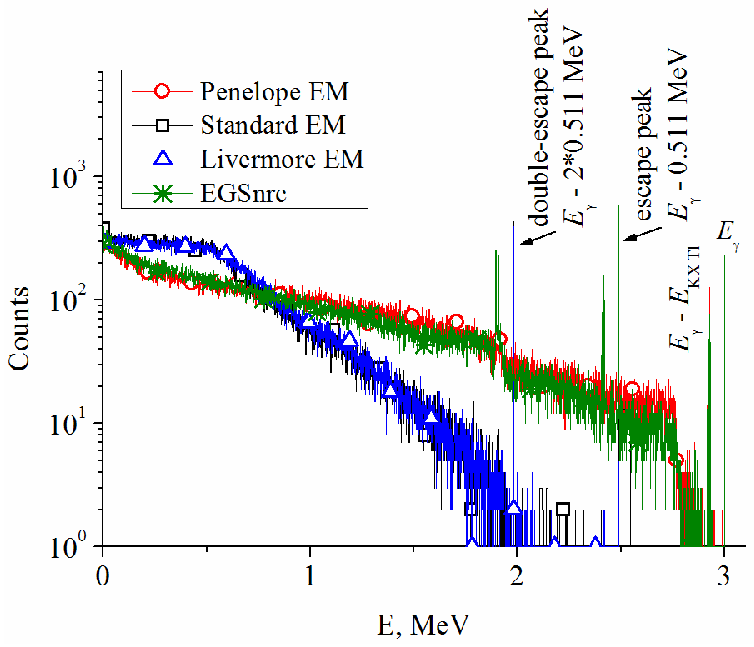}
\emph{\textbf{Fig.12.}} {\emph{The energy losses of gamma-quanta with the energy E$_{\gamma}$ = 3~MeV in TlBr-detector}}\\
\end{minipage}

\begin{center}
\emph{\textbf{Table 6.}} {\it The amount of event of the escape and double escape of annihilation gamma-quanta (TlBr),
E{$_{\gamma}$} = 3~ MeV}
\end{center}
\begin{center}
\begin{tabular}{|c|c|c|}
\hline
EM model & Escape & Double-escape\\
\hline
Standard & 13 & 435\\
\hline
Livermore & 5 & 395\\
\hline
Penelope & 235 & 5881\\
\hline
EGSnrc & 577 & 13433\\
\hline
\end{tabular}
\end{center}

\begin{center}
\textbf{\textsc{4. CONCLUSIONS}}
\end{center}
The comparison of the response functions of the HgI$_2$ and TlBr gamma-radiation detectors computed for all models of the electromagnetic interactions of the $Geant4$ package and $EGSnrc$ has been made in this work. In the energy region of gamma-quanta up to 0.4 MeV there is a good agreement between response functions of the investigated detectors calculated with using $EGSnrc$ and both $Livermore$ and $Penelope$ models of the $Geant4$ package. For wider gamma-quantum energy region between 0.026 and 3~MeV statistical parameters of the response functions simulated with $EGSnrc$ and $Geant4$ packages have a good agreement only when the $Penelope$ model of electromagnetic interactions has been used.

By the HgI$_2$-detector example it is shown that the comparison of the response functions received with using different simulation packages allow removing incorrect results and explaining the reason of their appearance.

\vspace{3mm}
\begin{center}

\end{center}
\end{multicols}
\end{document}